\documentclass[11pt]{article}

\usepackage{asp2006}
\usepackage{psfig}

\newcommand{\be}{\begin{equation}}
\newcommand{\ee}{\end{equation}}
\newcommand\eq{Equation}
\newcommand\eqs{Equations}
\newcommand\fig{Figure}

\newcommand{\kvec}{{\bf k}}

\begin{document}
\title{The Density of Coronal Null Points from Hinode and MDI}
\author{Dana Longcope}   
\affil{Dept. of Physics, Montana State Univ.,
  Bozeman, MT 59717, USA}    
\author{Clare Parnell}   
\affil{School of Mathematics and Statistics, 
University of St Andrews,\\St Andrews, Fife, UK, KY16 9SS}
\author{Craig De{F}orest}
\affil{Southwest Research Institute, 1050 Walnut Street, Suite 300\\
Boulder,CO 80302, USA}

\begin{abstract}
Magnetic null points can be
located numerically in a potential field extrapolation 
or their average density can be
estimated from the Fourier spectrum of a
magnetogram.  We use both methods to compute the null point density
from a quiet Sun magnetogram made with Hinode's NFI and from magnetograms from
SOHO's MDI in both its high-resolution and low-resolution modes.  
All estimates of the super-chromospheric column density 
($z>1.5$ Mm) agree with one another and with the
previous measurements: $3\times10^{-3}$ null points per square Mm of 
solar surface. 
\end{abstract}

\section{Null Point Density}

\begin{figure}[htb]
\centerline{\psfig{file=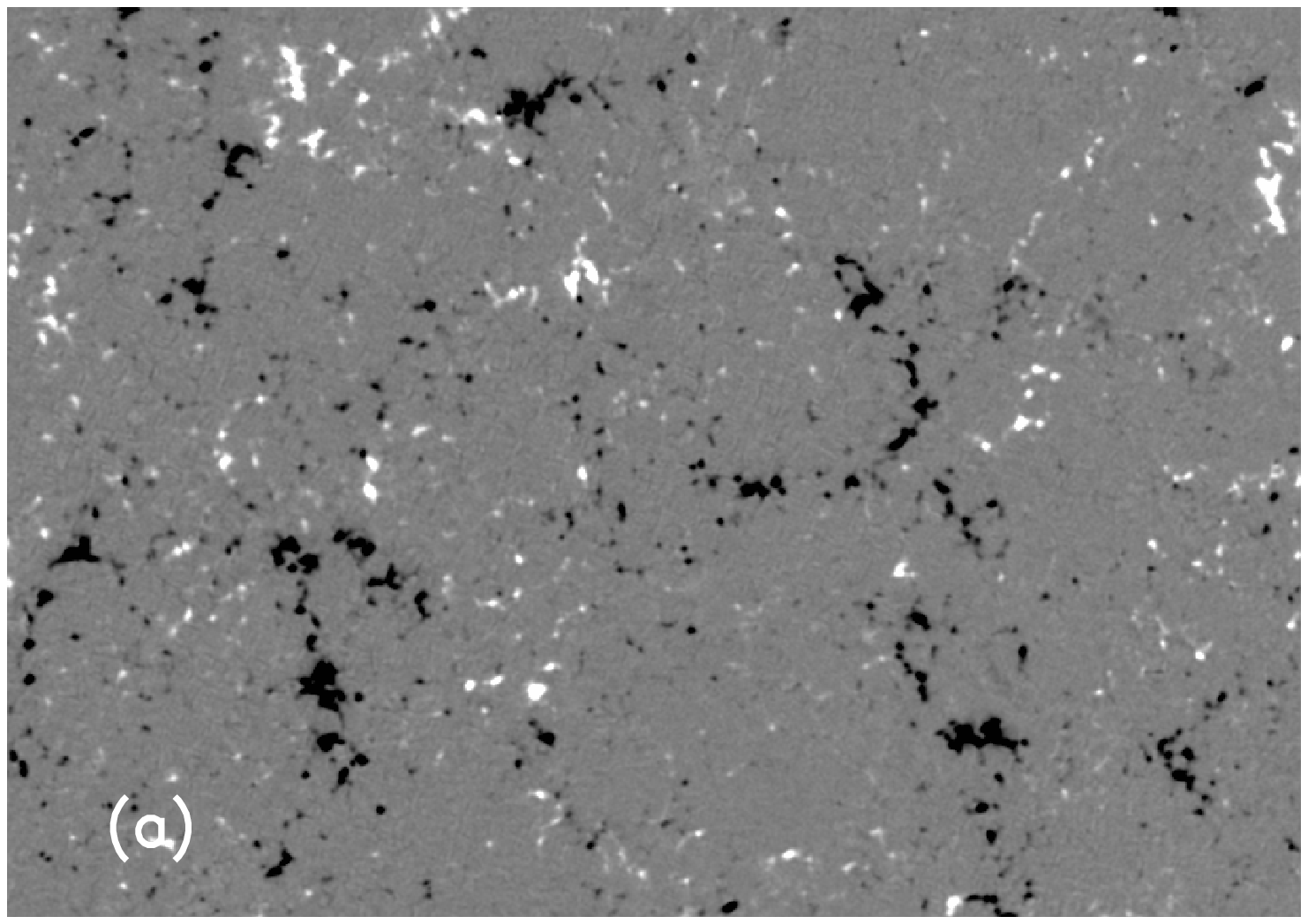,height=2.1in}
\psfig{file=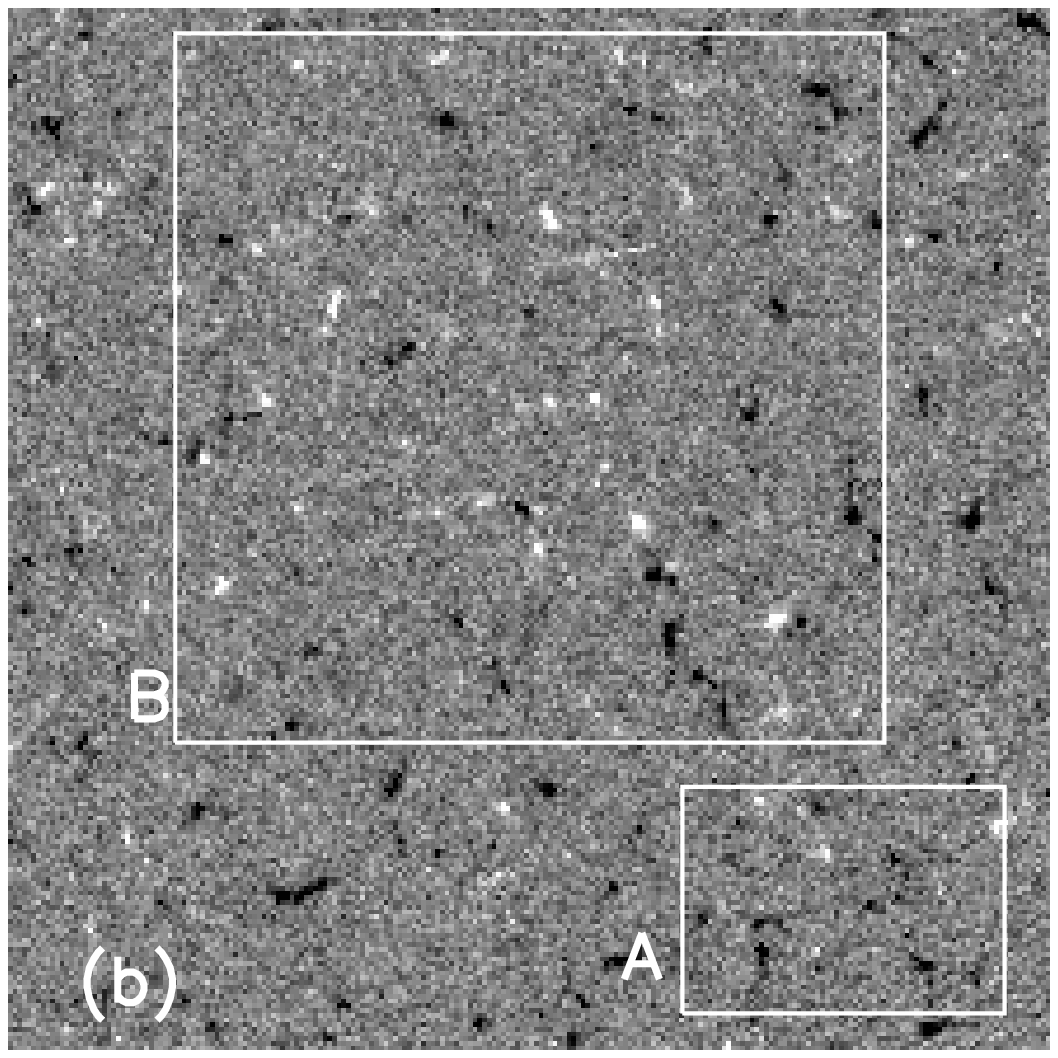,height=2.1in}}
\caption{Magnetograms of a section of quiet Sun from 19 Sept. 2008,
(scaled between $\pm100$ G).
(a) From the MDI NFI instrument.  (b) From SOHO/MDI in full-disk
mode.  Outlines show the field of view of (a) and of the
$300''\times300''$ images used below.}
	\label{fig:mgs}
\end{figure}

Line-of sight magnetograms, such as \fig\ \ref{fig:mgs}, show the
quiet sun to be composed of a disordered mixture of positive and
negative polarity.  The coronal magnetic field above is expected to be 
complex and vanish at a number of points called
null points \citep{Schrijver2002,Longcope2008e}.  These null points
might serve as a site of
heating \citep{Hassam1992,Craig1993} or of coronal jets
\citep{Shibata1992,Cirtain2007}.

The high resolution magnetogram from Hinode's NFI 
(\fig\ \ref{fig:mgs}a, with $0.16''\times0.16''$ pixels) 
reveals fine spatial structure, however, within a relatively
small field of view: $136''\times96''$.  
The MDI full-disk magnetogram (\fig\ \ref{fig:mgs}b) has lower 
resolution but from its larger field of view it is possible to extract
the same region (labeled {\sf A}) or a larger
$300''\times300''$ square centered at disk center ({\sf B}), similar
to those used by \citet{Longcope2008e}.

Potential fields are extrapolated from lower boundaries composed of
the NFI magnetogram (\fig\ \ref{fig:mgs}a) and sub-field {\sf A} of
the MDI magnetogram.  This region is near enough to disk center that
no compensation is made for surface curvature --- the line-of-sight
field is used as the vertical field, $B_z(x,y,0)$.   The extrapolation
yields field components on Cartesian grid points.  The
algorithm of \citet{Haynes2007} identifies each point at which
tri-linear interpolation of the gridded field vectors
vanish: magnetic null points.

\begin{figure}[hbt]
\centerline{\psfig{file=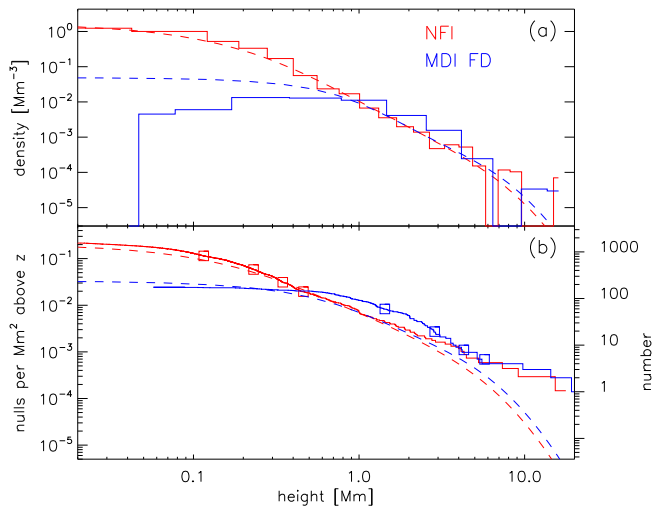,width=4.0in}}
\caption{The density of null points above the Hinode magnetogram
(red) and sub-field {\sf A} of the MDI full disk magnetogram (blue).  (a)
The density, $\rho_N(z)$, of null points found on a gridded extrapolation
(solid) and by spectral estimate (dashed).  (b) The column density,
$N_n(z)$ found by integrating $\rho_N(z)$ downward.  Left axis shows
the value in Mm$^{-2}$, the right axis gives the total number of null
points found on the grid.}
	\label{fig:cpp}
\end{figure}

The heights of null-points directly found this way 
can be summarized as a density,
$\rho_N(z)$, shown as the solid curves in
\fig\ \ref{fig:cpp}a.  The null column,
$N_z(z)$, shown in \fig\ \ref{fig:cpp}b
is the number of null points, per surface area, 
above a given height.
For the surface area of the entire magnetogram (the right axis) the
number of nulls above $z\simeq8$ Mm falls to only a handful, and the
cumulative density suffers from statistical errors.

Comparison reveals
that additional spatial resolution yields more null points in the potential
extrapolation from an ideal photospheric surface ($z=0$), but the additional
null points are, as expected, at very low altitude.  Hence, the number
of coronal null points ($z>1.5$ Mm) is much less dependent on resolution.

\section{The Spectral Estimate}

\begin{figure}[htb]
\centerline{\psfig{file=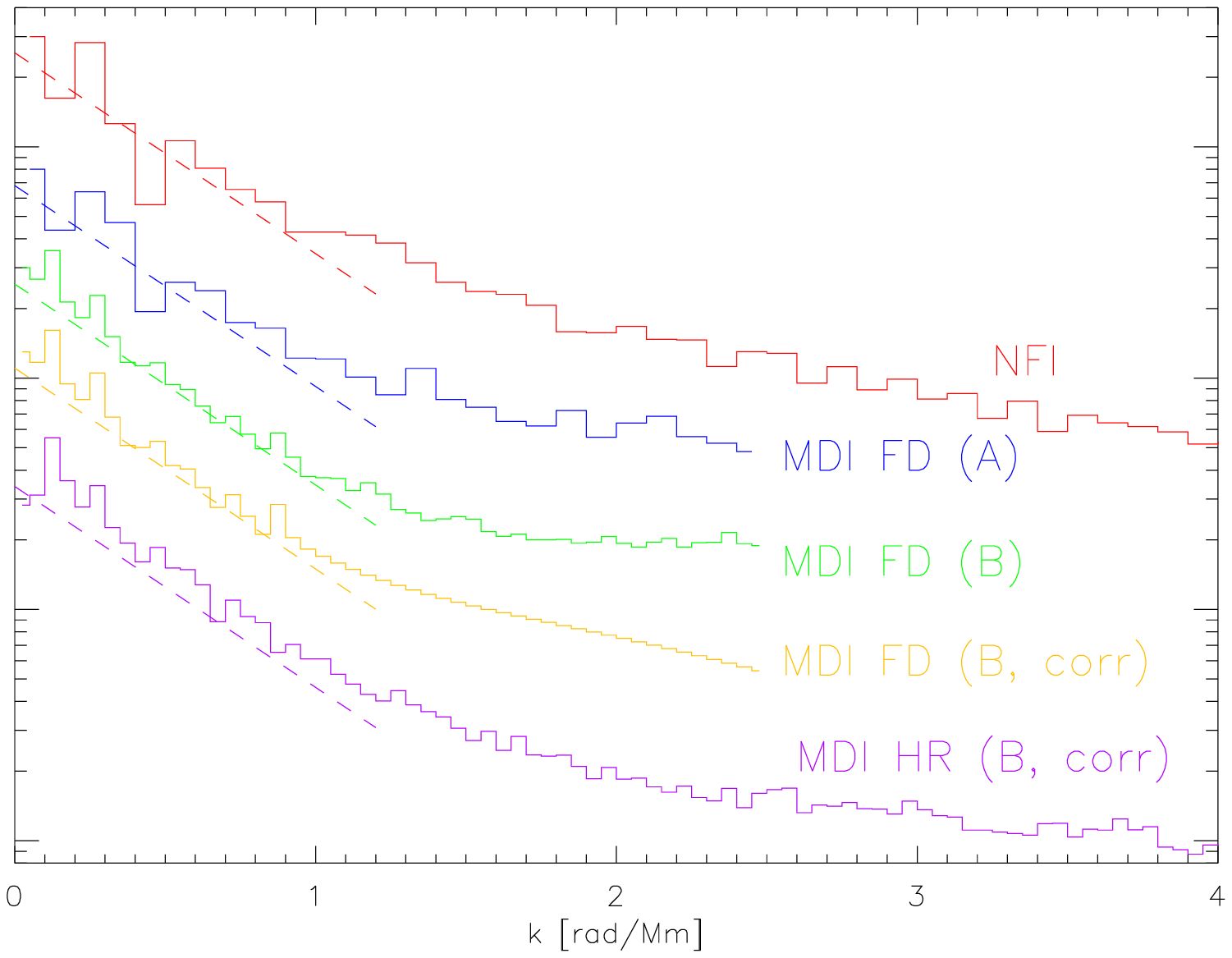,width=2.7in}%
\psfig{file=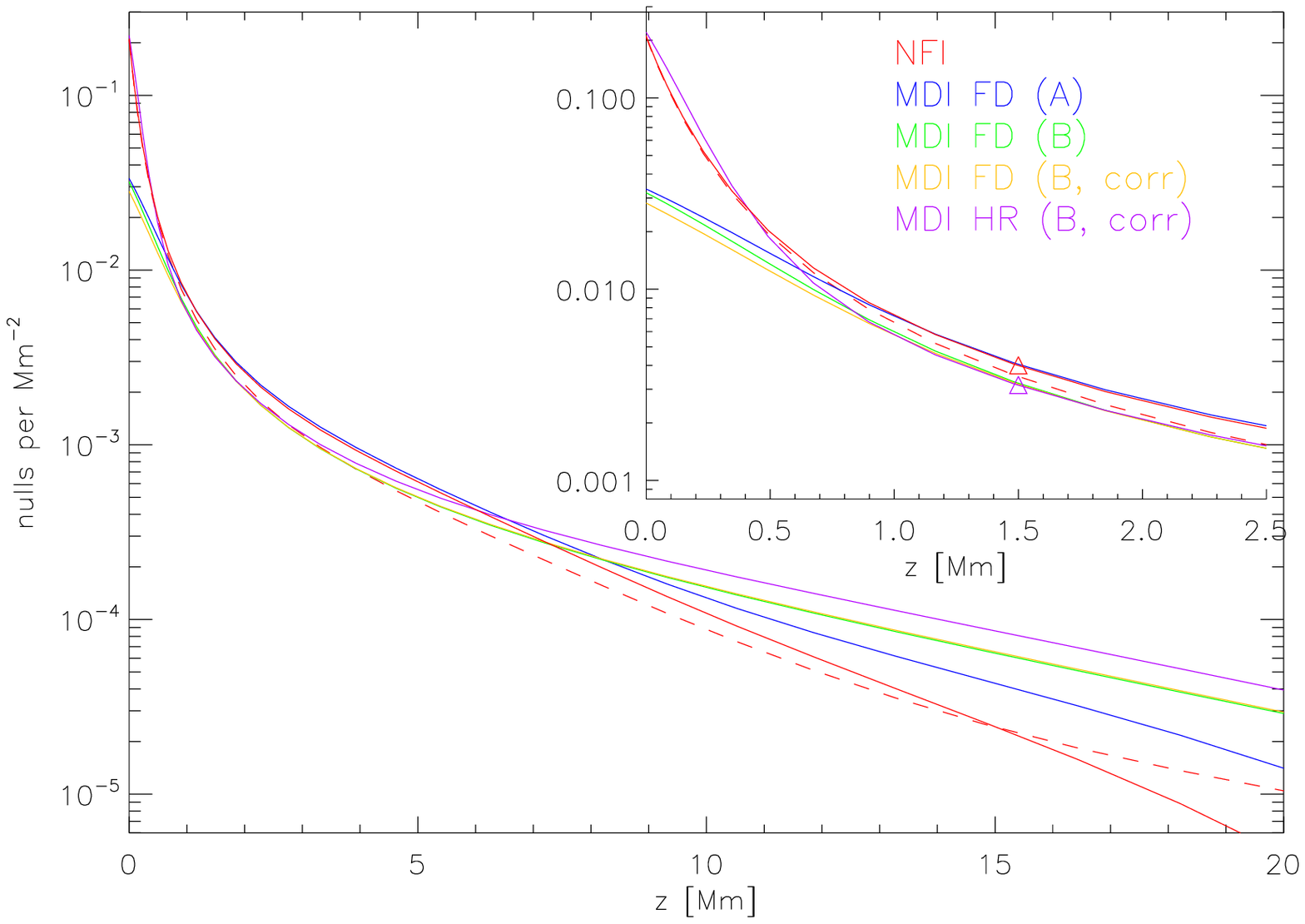,width=2.7in}}
\caption{(a) The isotropized spectra, $S(k)$, for different magnetograms,
plotted in different shades, offset vertically for clarity.  Dashed
lines are $S(k)\sim e^{-2kd}$, with $d=1.2$ Mm, for comparison.
(b) Spectral estimates of the null column densities
using the same shades.  Triangles mark the column above $z=1.5$ Mm.
The dashed  line is from the NFI spectrum, but with $\bar{B}_z=0$.}
	\label{fig:spec}
\end{figure}

An alternative to direct identification of null points is an estimate
based on the isotropized Fourier spectrum of the
magnetogram, $S(k)$, developed  by \citet{Longcope2003b}.
Figure \ref{fig:spec}a shows the spectra from various magnetograms, 
computed out to 
$k_c=2\sqrt{\pi}/\Delta x$ in order that the the total 
Fourier-space area, $\pi k_c^2$, matches
that occupied by the actual magnetogram with $\Delta x\times\Delta x$
pixels.

The spectral estimate of null density depends on the
variance, $\sigma_z^2(z)$, 
of the vertical component of the extrapolated field
as well as on an inverse characteristic length scale,
$q(z)$, defined by a related integral
\be
  \sigma_z^2(z) = 2\pi \int_0^{\infty} S(k)e^{-2kz}\, k\, dk ~~,~~
  q^2(z) = {2\pi \over \sigma_z^2(z)}
  \int_0^{\infty} S(k)e^{-2kz}\, k^3\, dk ~~.
        \label{eq:sig+q}
\ee
\citet{Longcope2003b} showed that under the assumption that
$B_z(x,y,0)$ was a homogeneous, Gaussian random field, the density of
magnetic null points is
\be
  \rho_N(z) ~=~ G(\bar{B}_z/\sigma_z)\, q^3(z) ~~,
        \label{eq:rho}
\ee
where $G(\zeta)$ is a dimensionless function \citep{Longcope2003b}.
The function depends on the mean vertical field strength $\bar{B}_z$ and also
weakly on another moment of the spectrum.
The spectral estimate, \eq\ (\ref{eq:rho}), 
is plotted in \fig\ \ref{fig:cpp}a, as a dashed curve for
each of the two magnetograms.  Integrating $\rho_N(z)$ downward from
infinity gives the column densities, $N_n(z)$, plotted as dashed
curves in \fig\ \ref{fig:cpp}b.  In the case of NFI it appears that the
spectral estimate is reasonably accurate over heights with more than a
handful of null points above.

The spectral estimate depends on the isotropized spectra primarily through
integrals in \eq\ (\ref{eq:sig+q}).  Different patches of the
quiet Sun, such as regions {\sf A} and {\sf B} of \fig\ \ref{fig:mgs},
have very similar spectra, as shown in \fig\ \ref{fig:spec}a.
Larger fields, such as {\sf B}, sample
$S(\kvec)$ more densely and thus yield 
more accurate spectra.  Higher resolution, on
the other hand allows spectra to extend to much higher wave numbers, as
does the NFI curve in \fig\ \ref{fig:spec}a.  The spectra from
instruments of all resolution and fields-of-view agree relatively well
over the lowest wave numbers, $k\le 1$ rad/Mm.

The moderate wave numbers of the MDI
spectra are affected by white noise and the modulation transfer
function of the imaging system, including an intentional defocusing
in full-disk mode \citep{Scherrer1995}.
Correcting for these effects
\citep{Longcope2008e} yields spectra even closer to those of NFI.
The NFI magnetogram is corrected by deconvolution with a PSF
calculated from the lunar transit of July 2008; this has a small but
noticeable effect in the data.

Using each of the spectra from \fig\ \ref{fig:spec}a in the spectral
estimate leads to column densities shown in \fig\ \ref{fig:spec}b.
The fact that all of these agree over $z\ge 700$ km follows from the
exponential factor in \eqs\ (\ref{eq:sig+q}).  Due to
this factor the spectral estimate at a height $z$ will depend on the
spectral range $k<1/z$, where all spectra are similar.  The fine-scale
structure contributing to the high wavenumbers in both the Hinode NFI
and the high-resolution MDI spectra create a layer of
null points close to the photosphere ($z<500$ km).
Departures over $z>10$ Mm are due to differences in $k<0.1$ rad/Mm,
where the different fields of view lead to different spectral forms.

Evaluating the columns just above the chromosphere, at $z=1.5$ Mm,
gives $N_n(1.5\,{\rm Mm})=3.9\times10^{-3}\,{\rm Mm}^{-2}$ for NFI and
$N_n(1.5\,{\rm Mm})=3.1\times10^{-3}\,{\rm Mm}^{-2}$ for MDI in high
resolution (triangles in the inset).  
The latter is almost exactly the value found in a sample
of 562 MDI quiet Sun magnetograms over 1996--1998 and 2006--2007
\citep{Longcope2008e}.  Both the NFI and MDI full-disk estimates from
field {\sf A} are greater over the range $1<z<2$ Mm (expanded in the
inset) primarily due to the mean field over the small field of
view, {\sf A}: $\bar{B}_z\simeq-2.5$ G.  The larger field
of view ({\bf B}) is much more flux-balanced 
($\bar{B}_z\simeq -0.28$ G). 
Using the NFI spectrum 
with $\bar{B}_z=0$ in \eq\ (\ref{eq:rho}) gives the estimates from the
dashed line, falling much closer to those from field {\sf B}.

\acknowledgements 

This work was supported by LWS TR\&T.

\bibliography{/Users/danalongcope/stuff//long_abbrevs.bib,/Users/danalongcope/stuff/full_lib.bib}

\end{document}